\documentclass[preprint2]{aastex}		

\received{11 February 2004}
\revised{}
\accepted{22 March 2004}

\cpright{AAS}{2004}

\shortauthors{Fynbo et al.}
\shorttitle{XRF~030723: afterglow and associated SN}

\slugcomment{Submitted to ApJ, February 2004}

\newcommand{\Msun}{~M_\odot}
\newcommand{\lsim}{\raise0.3ex\hbox{$<$}\kern-0.75em{\lower0.65ex\hbox{$\sim$}}}
\newcommand{\gsim}{\raise0.3ex\hbox{$>$}\kern-0.75em{\lower0.65ex\hbox{$\sim$}}}

\begin{document}

\title{On the Afterglow of the X-Ray Flash of July 23 2003: Photometric 
evidence for an off-axis Gamma-Ray Burst with an associated 
Supernova?\footnote{\rm Based on 
observations made with ESO 
Telescopes at the Paranal and La Silla Observatories under programme ID 
71.A-0355(A,B,C,F,G)}}

\author{
J.~P.~U.~Fynbo\altaffilmark{2,3},
J.~Sollerman\altaffilmark{4},
J.~Hjorth\altaffilmark{3},
F.~Grundahl\altaffilmark{2,3},
J.~Gorosabel\altaffilmark{5,6},
M.~Weidinger\altaffilmark{2,7},
P.~M\o ller\altaffilmark{7},
B.~L.~Jensen\altaffilmark{3},
P.~M.~Vreeswijk\altaffilmark{8},
C.~Fransson\altaffilmark{4},
E.~Ramirez-Ruiz\altaffilmark{9},
P.~Jakobsson\altaffilmark{3},
S.~F.~J{\o}rgensen\altaffilmark{3},
C.~Vinter\altaffilmark{3},
M.~I.~Andersen\altaffilmark{10},
J.~M.~Castro Cer\'on\altaffilmark{6},
A.~J.~Castro-Tirado\altaffilmark{5},
A.~S.~Fruchter\altaffilmark{6},
J.~Greiner\altaffilmark{11},
C.~Kouveliotou\altaffilmark{12},
A.~Levan\altaffilmark{13},
S.~Klose\altaffilmark{14},
N.~Masetti\altaffilmark{16},
H.~Pedersen\altaffilmark{3},
E.~Palazzi\altaffilmark{17},
E.~Pian\altaffilmark{16,17},
J.~Rhoads\altaffilmark{6},
E.~Rol\altaffilmark{15},
T.~Sekiguchi\altaffilmark{18},
N.~R.~Tanvir\altaffilmark{19},
P.~Tristram\altaffilmark{20},
A.~de Ugarte Postigo\altaffilmark{6},
R.~A.~M.~J.~Wijers\altaffilmark{15}, and
E.~van~den~Heuvel\altaffilmark{15}
}
\altaffiltext{2}{Department of Physics and Astronomy, University of Aarhus,
Ny Munkegade, DK--8000 \AA rhus C, Denmark}
\altaffiltext{3}{Niels Bohr Institute, Astronomical Observatory, University 
of Copenhagen, Juliane Maries Vej 30, DK--2100 Copenhagen~\O, Denmark}
\altaffiltext{4}{Stockholm Observatory, Department of Astronomy, AlbaNova, 
S-106 91 Stockholm, Sweden}
\altaffiltext{5}{Instituto de Astrof\'{\i}sica de Andaluc\'{\i}a (CSIC), 
c. Camino Bajo de Hu\'etor, 24, E-18.008 Granada, Spain}
\altaffiltext{6}{Space  Telescope Science  Institute, 3700  San Martin
Drive, Baltimore, MD 21218, USA}
\altaffiltext{7}{European Southern Observatory, Karl Schwarzschild-Strasse 2,
D-85748 Garching, Germany}
\altaffiltext{8}{European Southern Observatory, Alonso de C{\'o}rdova 3107, 
Casilla 19001, Santiago 19, Chile}
\altaffiltext{9}{School of Natural Sciences, Institute for Advanced Study, 
Einstein Drive, Princeton, NJ 08540, USA; Chandra Fellow}
\altaffiltext{10}{Astrophysikalisches Institut Potsdam, An der Sternwarte 16,
D-14482 Potsdam, Germany}
\altaffiltext{11}{Max-Planck-Institute for Extraterrestrial Physics,
D-85741 Garching, Germany}
\altaffiltext{12}{Universities Research Association, Marshall Space Flight
Center (NASA), Huntsville, AL 35812}
\altaffiltext{13}{X-ray Astronomy Group, Department of Physics and Astronomy,
Leicester University, Leicester LE1 7RH, UK}
\altaffiltext{14}{Th\"uringer Landessternwarte Tautenburg,
D--07778 Tautenburg, Germany}
\altaffiltext{15}{Astronomical Institute `Anton Pannekoek',
NL--1098 SJ Amsterdam, The Netherlands}
\altaffiltext{16}{IASF/CNR, Sezione di Bologna, Via Gobetti 101,
I--40129 Bologna, Italy}
\altaffiltext{17}{INAF, Osservatorio Astronomico di Trieste, via Tiepolo,
I--34131 Trieste, Italy}
\altaffiltext{18}{Solar Terrestrial Environment
Laboratory, Nagoya University, Nagoya 464-8601, Japan}
\altaffiltext{19}{Department of Physical Sciences, University of Hertfordshire,
College Lane, Hatfield, Hertfordshire AL10 9AB, UK}
\altaffiltext{20}{Canterbury University's Mt John Observatory,
Lake Tekapo, New Zealand}

\begin{abstract}
We present optical and near-infrared follow-up observations of the
X-Ray Flash (XRF) of July 23 2003. Our observations in the R-band cover
the temporal range from 4.2 h to 64 days after the high energy event. 
We also present the results of multicolor imaging extending to the 
K-band on three epochs. The light curve of the R-band afterglow the 
first week after the burst is similar to the light curve for long 
duration Gamma-Ray Bursts (GRBs), i.e.,\ a broken power-law with a late 
time slope of $\alpha\approx2.0$ ($F_{\nu} \propto t^{-\alpha}$). Furthermore,
the spectral energy distribution (SED) has a power-law 
($F_{\nu} \propto \nu ^{-\beta}$) shape with slope ${\beta}\approx1.0$. 
However, the decay slope at $t<1$ day is shallow, consistent with
zero. This is in qualitative agreement with the prediction that XRFs are 
off-axis classical GRBs. After the first week there is a strong bump in 
the light curve, which peaks at around 16 days. 
The SED after the peak becomes 
significantly redder. We discuss the possible interpretations of this bump, 
and conclude that an underlying supernova is the most likely
explanation since no other model appears consistent with the evolution of the 
SED. 
Finally, we present deep spectroscopy of the burst both in the afterglow and 
in the bump phase. A firm upper limit of $z=2.3$ is placed on the redshift
of XRF~030723 from the lack of Ly$\alpha$ forest lines in the spectrum
of the afterglow. The lack of significant absorption and emission lines 
in either of the two spectra excludes a spectroscopic redshift 
determination. 
\end{abstract}


\section{INTRODUCTION}

X-Ray Flashes (XRFs) are transient sources of X-ray photons
which are distributed iso\-tro\-pi\-cal\-ly 
on the sky. Their existence as a
class with a different nature than X-Ray Bursters in the 
Galaxy was first put forward by Heise et al.\ (2001) based on 
data from the BeppoSAX satellite. 
They introduced an operational definition for XRFs, namely a 
fast transient source with duration less than 1000~s in the
Wide Field Camera (covering 2--25~keV) that did not trigger 
the Gamma Ray Burst Monitor (covering 40--700~keV). 
BeppoSAX detected 20 such events in the 6 years it was operational
(Heise, in 't Zand, \& Kippen 2003). 
A HETE-2 burst is classified as an XRF if its X-ray fluence ($S$) exceeds its 
gamma-ray fluence, i.e., 
$\log [ S($2--30~keV)/$S$(30--400~keV)$] >0$
(Lamb, Donaghy, \& Graziani 2003). Classical Gamma-Ray Bursts (GRBs) 
are defined as having 
$\log [ S($2--30~keV)/$S$(30--400~keV)$] < -0.5$, 
while bursts belonging to the intermediate class are classified as 
X-ray rich GRBs, following Castro-Tirado et al.\ (1994).

XRFs can be interpreted as the same phenomenon as classical 
GRBs with the difference being that the high energy 
spectra for XRFs are softer than for GRBs (Heise et al.\ 2001; 
Barraud et al.\ 2003). The high energy 
spectra ($\nu F_{\nu}$) of GRBs are well described by the so 
called Band function (Band et al.\ 1993), which is composed of two 
smoothly connected 
power-laws. The energy at which the two power-laws connect is referred 
to as $E_\mathrm{peak}$ and is where most of the energy usually is
emitted. For classical GRBs $E_\mathrm{peak}$ is typically
a few 100~keV \cite{preece00}. The high energy spectra of XRFs 
are also well fitted by the Band function, but with values 
of $E_\mathrm{peak}$ below 100~keV and in some cases even below 
10~keV \citep{kippen02,barraud03}. For classical GRBs there is a 
correlation between spectral hardness, defined as the fluence in the X-ray
band divided by the fluence in the gamma-ray band, and the 
fluence in the gamma-ray band in the sense that hard bursts
have a higher total fluence \cite{nemiroff94}. XRFs follow 
this correlation 
\cite{barraud03}. These properties of GRBs and XRFs would be 
naturally explained if XRFs were GRBs at very large redshifts.
However, the lack of excessive time dilation compared
to GRBs argues against this interpretation \cite{heise01}. Moreover,
Amati et al.\ (2002) find in a sample of 12 BeppoSAX GRBs  
with known redshifts that the total isotropic 
equivalent energy radiated in the 1--10000~keV range is positively correlated 
with $E_\mathrm{peak}$. This also argues against very large redshifts
for XRFs. 
Instead the current view is that XRFs are either {\it i)} the result of
classical GRBs seen off-axis (Yamazaki et al.\ 2002, 2003; Dado et al.\
2003; Rhoads 2003), {\it ii)} so-called dirty fireballs, which are
relativistic jets with a larger baryon load and hence (assuming external
shocks) lower $\Gamma$-factor than those of classical GRBs (Dermer,
Chiang, \& B{\"o}ttcher 1999; Heise et al.\ 2001), or {\it iii)} fireballs
with large $\Gamma$-factors and/or low baryon loading that in the case of
internal shocks lead to the emission of less energetic photons (Zhang \&
M{\'e}sz{\'a}ros 2002; Barraud et al.\ 2003).

Prior to XRF~030723, four XRFs had been localized to arcmin accuracy. 
The first was the BeppoSAX burst XRF~011030 (e.g., Heise et al.~2001) for 
which a likely afterglow was detected at radio (Taylor et al.~2001) and 
X-ray wavelengths (Harrison et al.~2001). The first XRF localized by 
HETE-2 (XRF~011130) did not lead to an unambiguous afterglow detection 
despite extensive efforts at optical, X-ray and radio wavelengths.
XRF~020427, localized by BeppoSAX, exhibited an X-ray afterglow,
localized to arcsec accuracy by 
the Chandra X-ray Observatory (Fox 2002). The first HETE-2 XRF for 
which an afterglow was identified was XRF~020903 (Ricker et al.~2002). 
A candidate optical and radio afterglow was detected and a
bright galaxy coincident with the transient was found at $z = 0.251$ 
(Soderberg et al.~2003). 
Finally, Watson et al.\ (2004) suggest that the Integral burst GRB~031203 was 
actually an XRF based on the modeling of a dust-echo in our galaxy observed 
by XMM (Vaughan et al.\ 2004). For GRB~031203 a radio afterglow 
(Frail 2003), but no optical afterglow, has been detected. The likely redshift 
of this burst is only 0.105 based on the spectrum of a galaxy coincident with 
the position of the radio afterglow (Prochaska et al.~2004).

HST observations have been obtained of the fields of XRF 011030, 020903, 
and 020427. In all cases a candidate host galaxy has been identified. 
XRFs 011030 and 020427 exhibited blue $\mathrm{R}\sim 24$ host galaxies, 
typical of GRB host galaxies (Bloom et al.~2003). The galaxies are 
probably not at a very high redshift ($z\leq3.5$; Bloom et al.~2003). 
The host 
galaxy of XRF~020903 exhibits a complex morphology (Levan et al.\ 2002; 
Soderberg et al.~2003). A single emission line at 8485 {\AA} has been 
detected from the host galaxy of XRF~020427 (van Dokkum \& Bloom 2003).

XRF~030723 was detected by the FREGATE, WXM, and SXC 
instruments on-board the HETE-2 satellite on July 23.26965 2003 UTC 
(HETE trigger H2777). The event was localized with the SXC 
to a 2 arcmin radius error circle at high galactic latitude 
(b = 50$^\mathrm{o}$) in the constellation Pisces Aus \cite{prigozhin03}.
The burst duration (T$_{90}$) was 25 s. The total fluence in the
7--30~keV band was $\sim$2$\times$10$^{-7}$ ergs cm$^{-2}$ and in the
30--400~keV band the fluence was less than 7$\times$10$^{-8}$ ergs 
cm$^{-2}$. It was hence clearly an XRF according to the HETE-II 
definition.  
Observations of the X-ray afterglow to XRF~030723 has been reported
by Butler et al.\ (2003). In the radio band only an upper limit of 
180 $\mu$Jy at 8.46 GHz (July 26.42 UT) has been reported (Soderberg, 
Berger, \& Frail 2003). In this paper we present a comprehensive optical 
and near-infrared (near-IR) study of its afterglow - the first such study 
for any XRF.

\section{OBSERVATIONS}
 
We initiated optical follow-up observations of the SXC error-box
4.2 h after the flash. This was done near morning twilight from the 
Danish 1.5-m Telescope (D1.5m) at ESO's La Silla Observatory. Based 
on these observations we flagged the object later found to be 
the optical afterglow as the brightest source ($R=20.9$) not present 
in the DSS. Images were secured 
at the D1.5m also on the following night to facilitate detection of
the afterglow as a transient source. Comparison of the images from 
the first and second epoch observations revealed no apparent transients 
in the error-box \cite{jensen03}. This later turned out to be due to the 
small initial variability of the afterglow: 
within the errors the afterglow had the same R-band 
magnitude 4.2~h and 21.0~h after the burst. The detection of the optical 
afterglow was subsequently reported by Fox et al.\ (2003). The fact that we had observed
an initially flat light curve  
suggested to us that XRF~030723 was an off-axis GRB. Off-axis bursts should 
only be observable to modest redshifts (Yamazaki et al.\ 2002) and we 
therefore decided to conduct a targeted effort at detecting 
emission from the expected associated supernova (Stanek et al.\ 2003; Hjorth
et al.\ 2003a). We continued to observe the afterglow at
optical and near-IR wavelengths for the following 64~days using 
the D1.5m and ESO telescopes at the La Silla and Paranal 
observatories. The afterglow was also observed at 8.1~h after the
burst with the 0.6-m telescope on Mt. John in New Zealand in
wide MOA (Microlensing Observations in Astrophysics) 
RI- and BV-band filters.  
The full journal of observations is given in Table~\ref{tbl:journal}.
The data were reduced using standard techniques for de-biasing and 
flat-fielding.

\section{RESULTS}
\subsection{Astrometry}
We have determined the celestial position of the optical afterglow 
by comparison with 9 stars from the 2MASS catalogue. We find the 
position RA(2000) = 21:49:24.42, Dec(2000) = $-$27:42:47.30 within an
RMS scatter of 0.06 arcsec. The 2MASS astrometry is tied to the 
International Celestial Reference System (ICRS) via the Tycho 2 Catalog
and is accurate to 70-80 mas. The position we derive is consistent with 
the position reported by Fox et al.\ (2003). 
 
\subsection{The Light curve}
The optical photometry of the XRF was carried out using the 
{\tt DAOPHOT/ALLSTAR/ALLFRAME} photometry packages developed by Stetson 
(1987, 1994). Initially each image was run through {\tt DAOPHOT} and 
{\tt ALLSTAR} to produce a point-spread function (PSF) and a star 
list. Subsequently {\tt DAOMASTER} and {\tt DAOMATCH} were used on 
the {\tt ALLSTAR} photometry to derive 
positional transformations between the images and generate 
a master star list. From the master star list 7 isolated objects 
were selected as PSF stars and 
then a new PSF was generated for each image using these stars. 
After this step, we obtained the final PSF photometry using the 
{\tt ALLFRAME} program (Stetson 1994), with the PSFs and positional 
transformations generated above. The errorbars reported on the 
photometry are those produced by {\tt ALLFRAME}. The relative magnitudes
were transformed to the standard system using observations of the 
Mark A field. For the 
near-IR data we used aperture photometry and used 2MASS stars in the field
for photometric calibration. The zeropoint uncertainties are
of the order 0.03 mag for both the optical and near-IR bands.

The afterglow is not detected in the early MOA RI-band image, but it 
is clearly detected in the BV-image, which has a much lower 
sky background. We estimate an R-band magnitude by 
calibrating to internal V reference stars and assuming $\mathrm{V}-\mathrm{R} 
= 0.5$, the color determined on July 24.3. 

\begin{figure*}
\begin{flushleft}
{\includegraphics[width=2.00\columnwidth,angle=0,clip]{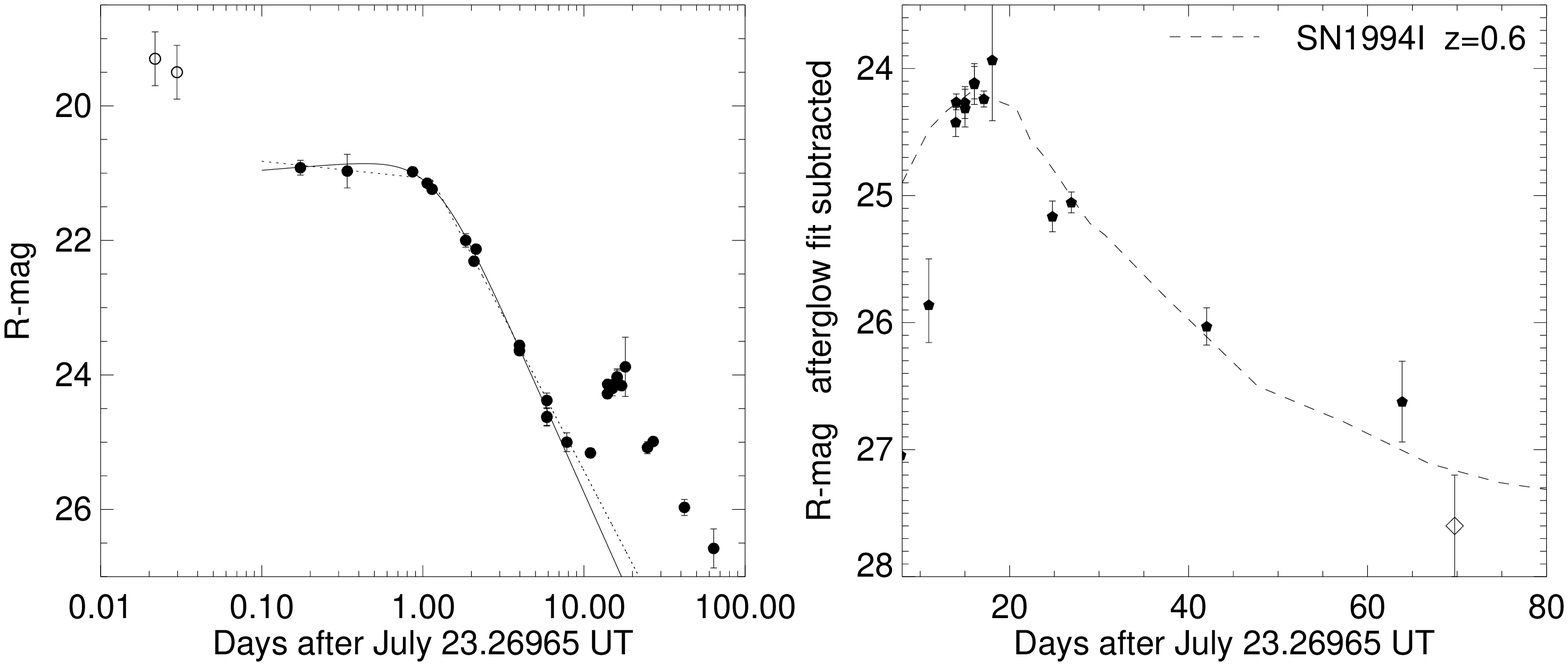}}
\caption{}{
{\it Left panel:} The R-band light curve of the afterglow of XRF~030723. 
The filled circles are the 
measurements from this work (Table~\ref{tbl:journal}) and the two open 
circles are the tentative 
early detections from the ROTSE-III telescope (Smith et al. 2003a). 
The dashed and solid lines are the result of broken power-law 
and Beuermann function fit to the first 12 data points.
{\it Right 
panel:} The late time bump. In this plot we have subtracted the 
extrapolation of the power-law component from the afterglow (based on the
Beuermann fit). The errors include the photometric error as well as the 
formal error from the subtraction of the Beuermann fit. 
We also include the latest reported detection (open circle) of Kawai et 
al.\ (2003).
The dashed curve shows the B-band light curve of the type Ic SN~1994I 
redshifted to $z=0.6$, and scaled up in flux by one magnitude. This 
combination of redshift and a very fast light curve provides the best
match to the late light curve.
}
\label{fig:lc}
\end{flushleft}
\end{figure*}

In the left panel of Fig.~\ref{fig:lc} we have plotted the R-band 
light curve ranging from 4.2~h to 64~days after the XRF. Here we also show
the tentative early un-filtered detections (2.7$\sigma$ and 3.1$\sigma$) 
from the ROTSE-III telescope (Smith, Akerlof, \& Quimby 2003a). The decay 
curve based 
on our data is consistent with being flat during the first 24~hr
after the burst. Assuming that the ROTSE-III detections are real, they imply a
declining phase during the first few hours. Around 1 day after the burst 
the decay slope steepens to about 
$\alpha = 2$ ($F_{\nu} \propto t^{-\alpha}$) and remains so for the 
following 4--5 days. About a week after the burst the light curve
starts to deviate from its fast decay. It then quickly 
rises to a secondary maximum peaked at around 16 days,
followed 
by another steep decline with power-law slope similar to that prior to the bump.

To quantify the properties of the early R-band light curve 
we have fit a broken power-law
\begin{equation}
\label{EQUATION:broken_power_law}
f_{R}(t) = \left \{
        \begin{array}{lll}
                f_{R}(t_b) {\left(\frac{t}{t_b}\right)}^{-\alpha_1}, &
                \mathrm{if}  &  t \le t_b \\
                f_{R}(t_b) {\left(\frac{t}{t_b}\right)}^{-\alpha_2}, &
                \mathrm{if}  &  t \ge t_b,
        \end{array}
             \right.
\nonumber
\end{equation}
to the first 12 data points, up to 6 days past the burst. 
We find $\alpha_1 = 0.10\pm0.06$, 
$\alpha_2 = 1.84\pm0.04$, $t_b=1.14\pm0.04$ days and $\chi^2 = 37$
for 7 degrees of freedom. Hence this functional form is formally 
rejected by the data. We then followed Beuermann et al.\ (1999) 
and fit an empirical function of the form
\[f_{R}(t) = [(k_1~t^{-\alpha_1})^{-n} + (k_2~t^{-\alpha_2})^{-n}]^{-1/n},\]
where $t$ is the time since the XRF, measured in days. For large values 
of $n$ this function approaches the broken power-law, whereas small values 
provide a smoothly broken
power-law. We found that $n\approx1.5$ gave the best fit. The decay 
slopes for $n=1.5$ are $\alpha_1 = -0.08\pm0.08$ and $\alpha_2 = 
2.15\pm0.08$ with $\chi^2=12$ for 7 degrees of freedom. In 
Fig.~\ref{fig:lc} we have over-plotted the two fits as a dotted 
(broken power-law) and a solid (Beuermann) line.

\subsection{Spectral Energy Distribution}
\label{SED}
The multiband observations of XRF 030723 allowed us to construct the
spectral energy distribution
(SED) at five epochs (July 24.4, July 25.4, July 27.2, Aug 6.3, and
Aug 19.0 UT). The optical and near-IR magnitudes were corrected for
Galactic reddening (E(B-V)=0.053, Schlegel, Finkbeiner, \& Davis 1998) and
transformed to specific flux using the conversion factors given
by Fukugita, Shimasaku, \& Ichikawa (1995) and Allen (2000), respectively. 
Given that
all multiband observations were not performed exactly at the above
mentioned five epochs, their corresponding fluxes were rescaled
assuming a power-law decay ($F_{\nu} \propto t^{-\alpha}$). Based on the
properties of the light curve, the assumed values for
$\alpha$ were $\alpha=2$ for July 24.4, July 25.4, July 27.2, Aug
19.0 UT and $\alpha=0$ for Aug 6.3 UT. In any case, the
near-IR/optical magnitudes are well clustered around the aforementioned
five epochs, so the final results are not strongly dependent on the
assumed values of $\alpha$. A power-law fit in the form $F_{\nu} \propto
\nu ^{-\beta}$ was carried out for the five epochs. The results are
summarized in Table~\ref{tbl:sed}.

A power-law provides a tolerable fit for the first four
epochs. The spectral index is consistent with being constant with a value 
around $\beta \sim 1.0$. There is no indication of a significant spectral 
bend due 
to extinction as seen for, e.g., GRB~000301C (Jensen et al.\ 2001),
GRB~000926 (Fynbo et al.\ 2001a; Price et al.\ 2001), or GRB~021004
(Holland et al.\ 2003). To constrain the amount of extinction we assumed
an intrinsic power-law shape of the SED, an SMC-like extinction curve 
from Pei (1992), and a redshift in the
range $z=0.3-1.0$ (see Sect.~\ref{redshift}). For $z=0.3$ the 
2$\sigma$ upper limit on the rest-frame $A_V$ is 0.5 mag and for $z=1.0$ 
the limit is 0.4 mag. 

\begin{figure*}[h]
\begin{flushleft}
{\includegraphics[width=2.00\columnwidth,angle=0,clip]{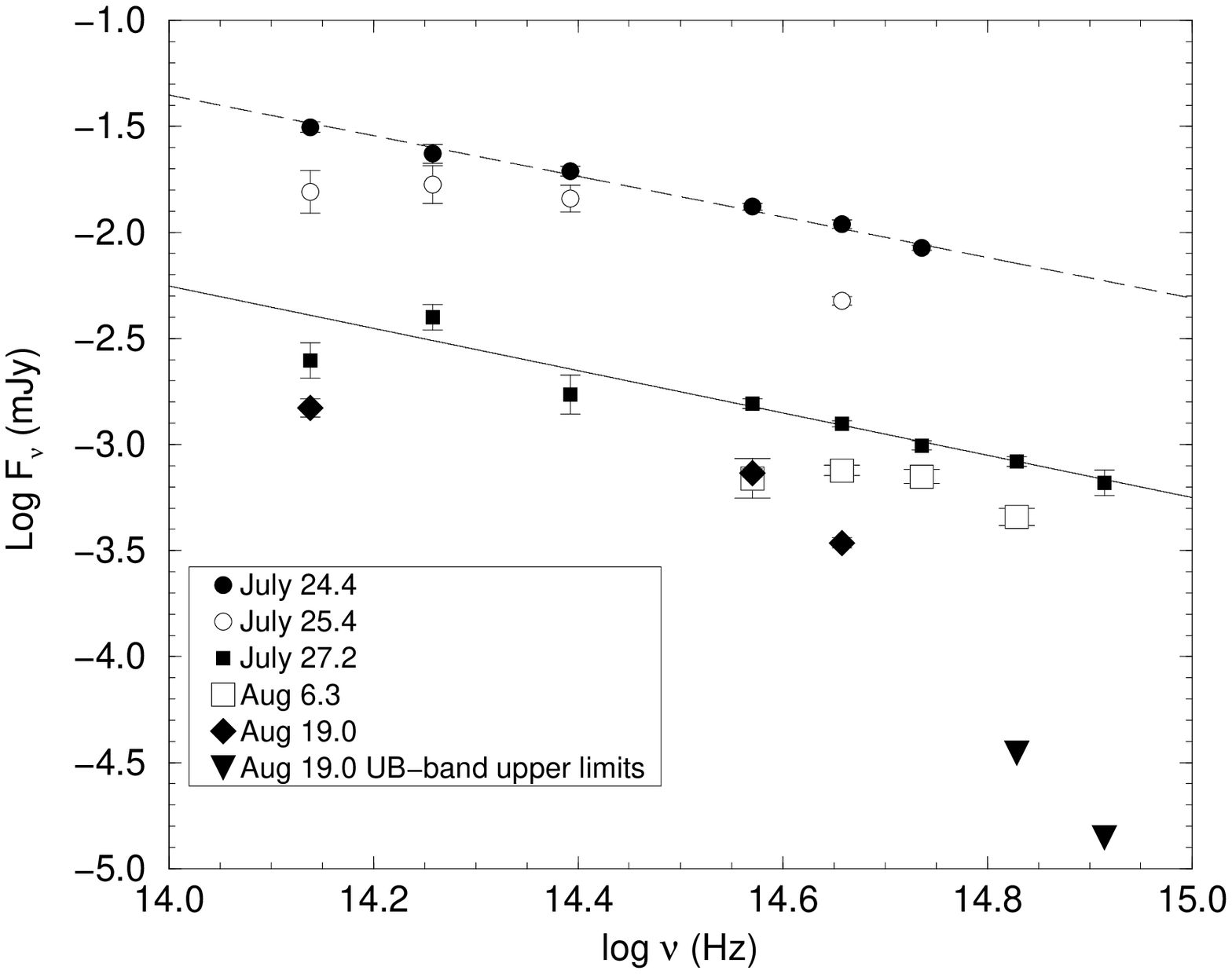}}
\caption{}{
The evolution of the SED from July 24.4 to Aug. 19.0. As seen, the SED is
consistent with a power-law in the first four epochs. However, after the
bump in the light curve there is a strong deviation away from the 
power-law SED due to a faster drop in the bluest bands. 
}
\label{fig:sed}
\end{flushleft}
\end{figure*}

Unlike the earlier epochs, the Aug 19.0 UT SED shows a clear deviation 
from the power-law ($\chi^2/\mathrm{dof} = 58.3$). Fig.~\ref{fig:sed} shows the 
five epoch SEDs. The constraining U- and B-band upper limits of 
Aug 19.0 UT (see filled triangles of Fig.~\ref{fig:sed}) indicate a clear 
deficit of flux at wavelengths shorter than $\sim$5000 \AA \ 
(observer frame) at this epoch.

\subsection{Spectroscopy}
\begin{figure*}[h]
\begin{flushleft}
{\includegraphics[width=2.00\columnwidth,angle=0,clip]{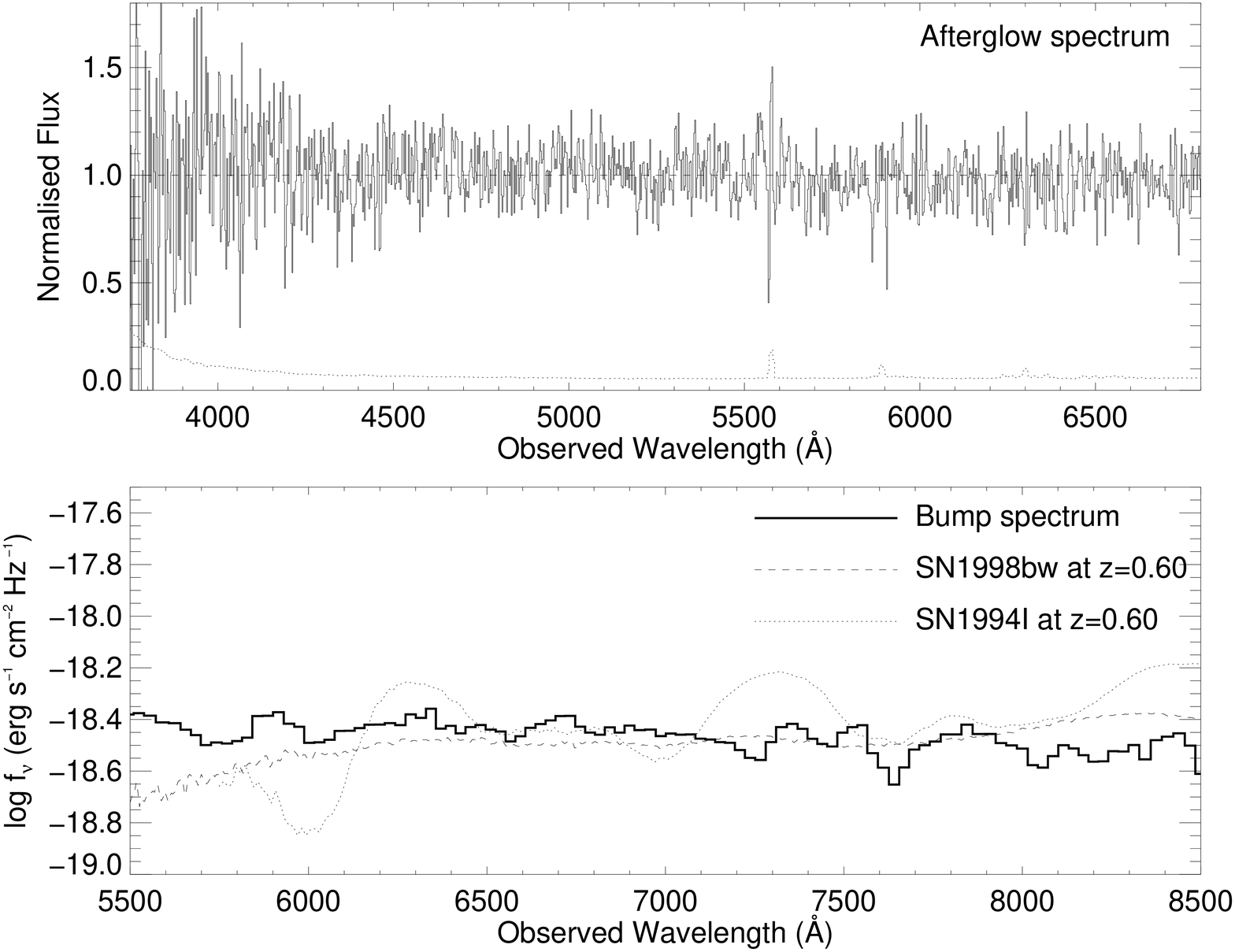}}
\caption{}{
{\it Upper panel:} The normalised spectrum of the optical afterglow from
July 26.3. The dotted line shows the noise spectrum. There are no
absorption lines significant at more than 2$\sigma$. We show only the 
spectral region from 3750~\AA\ to 6800 {\AA} that is largely unaffected 
by strong sky-subtraction residuals. 
{\it Lower panel:} The solid line shows the spectrum of XRF~030723 taken
during the peak of the light curve bump (Aug. 7.4+8.4). The spectrum
has been rebinned in 30 {\AA} wide wavelength bins. 
For comparison we show the spectra of SN~1998bw
(taken at 3 days past B-band peak, Patat et al.\ 2001) and SN~1994I 
(taken 1 day past B-band peak, Clocchiatti et al.\ 1996). 
Both spectra have been redshifted to $z=0.6$ (cf. Sect.~\ref{bumpnature}) and 
scaled in flux to match the level of the bump spectrum. The spectrum of the
bump is compatible with that of a SN with a high expansion velocity
similar to SN~1998bw or SN~2002ap, while we can probably exclude a SN with 
a low expansion velocity (as SN~1994I). 
}
\label{fig:spec1}
\end{flushleft}
\end{figure*}

Spectra of the source were obtained with the VLT on three epochs (see
Table~\ref{tbl:journal}). A 13.2~ks 
spectrum was obtained on July 26.3 when the afterglow had a magnitude
of V$\approx$23.3. We used the 300V grism with the GG375 order 
separation filter and a 1.0 arcsec wide slit providing a resolution of 
13.3 \AA \ over the spectral region from 3800~\AA \ to 8900~\AA. The 
spectrum shows a featureless continuum with no strong (larger than 
3$\sigma$) absorption
or emission lines superimposed. 
The normalized spectrum is shown in the upper panel of Fig.~\ref{fig:spec1}. 

We also secured spectra of the 
afterglow at the phase of the light curve bump. 
This was complicated by the bright moon, which was located near 
the field of XRF~030723 on the sky around the time of the bump.
The spectra were secured on Aug 7.4 and Aug 8.4 with a total
integration time of 13.5~ks. These spectra were obtained with 
the 600RI grism and the GG435 order separation filter and
a 1.0 arcsec wide slit providing a spectral coverage of 5300--8600 
\AA \ with a resolution of 6.6 \AA. The spectrum shows a faint 
continuum with no narrow absorption or emission lines detected
above 3$\sigma$ significance (Fig.~\ref{fig:spec1}). This spectrum 
is consistent with the broad band photometry from Aug. 6.3. 

\subsection{The Host Galaxy}
\begin{figure*}[h]
\begin{flushleft}
{\includegraphics[width=2.00\columnwidth,angle=0,clip]{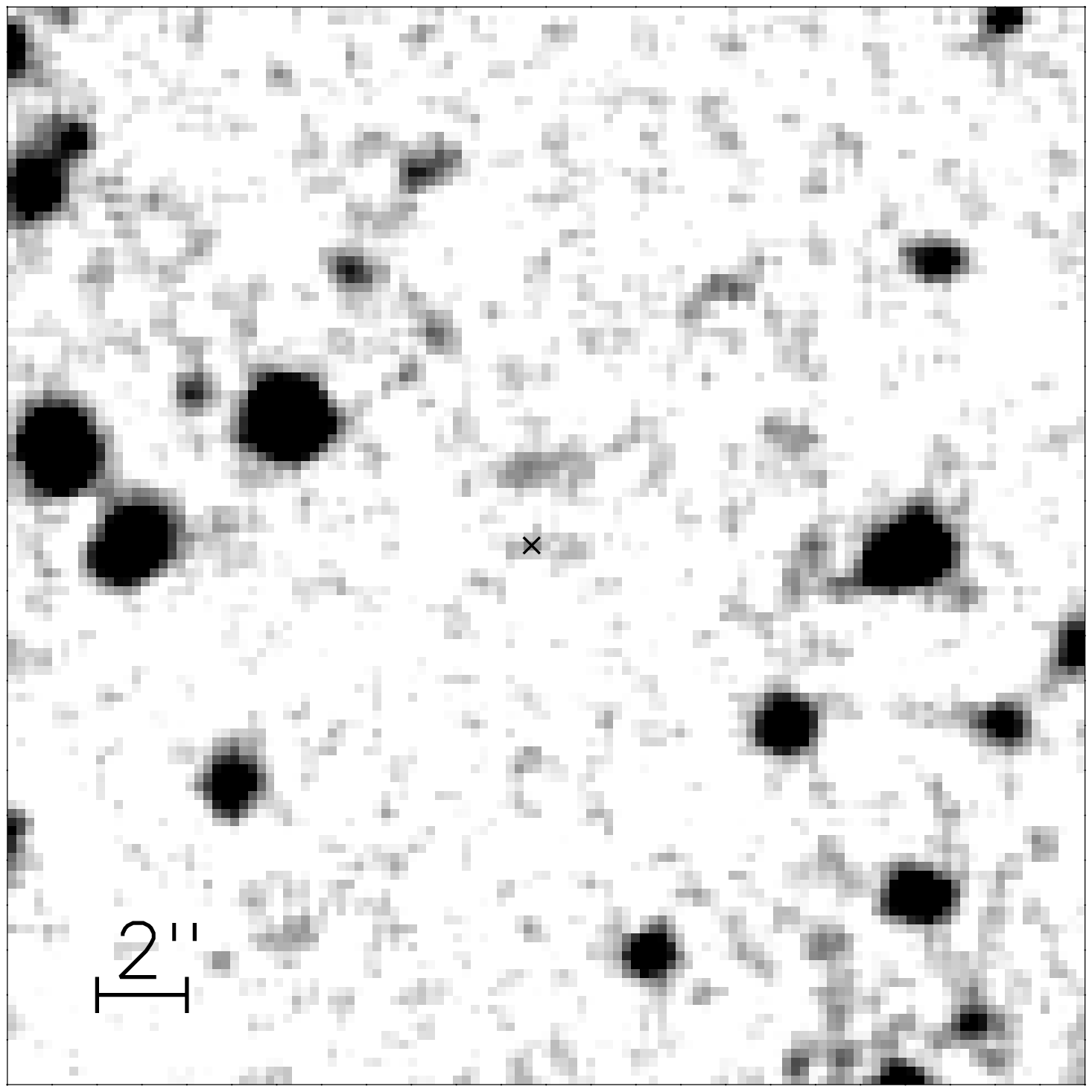}}
\caption{}{
The 24$\times$24 arcsec$^2$ field around the position of 
XRF~030723 from our latest VLT R-band image 64 days after the burst.
East is to the left and North is up.
The position of the afterglow is marked with a cross (the size
of the cross corresponds to the 10$\sigma$ error-circle on the 
astrometry). There is some evidence for an extended source with
position angle about 90$^\mathrm{o}$ (EofN) at the position of the 
afterglow (Fynbo et al.\ 2003). However, a later and deeper SUBARU 
image shows only a fainter point-source at the afterglow position
(Kawai et al.\ 2003). Further deep imaging is required to reveal 
the nature of the host galaxy.
}
\label{fig:host}
\end{flushleft}
\end{figure*}

In our latest R-band image from Sep. 25.1 there is 
a faint, extended source with magnitude $\mathrm{R}=26.8\pm0.4$ at the position
of the afterglow (Fig.~\ref{fig:host} and Fynbo et al.\ 2003). However, a 
deeper and later image taken with the SUBARU telescope shows the presence of a
$\mathrm{R}=27.6\pm0.4$ point source at the afterglow position (Kawai et al.\ 
2003). 
The nearest other galaxy is located about 2 arcsec north of the position of 
the afterglow. Whereas we cannot exclude this object as the host galaxy 
it is most likely unrelated to XRF~030723. We therefore consider 
$\mathrm{R}\approx27$ to be an upper limit on the magnitude of the host galaxy 
of XRF~030723. Further deep imaging is required to reveal the nature
of the host galaxy.

\section{DISCUSSION}

\subsection{The Afterglow}

This study of XRF~030723 is the first example of an XRF optical afterglow
that has been extensively monitored. The properties of the early light curve 
and SED are consistent with the hypothesis
that XRFs are due to the same phenomenon as classical GRBs, but seen at a
larger viewing angle. In this
scenario the early light curve is the result of the competition between 
two effects: {\it i)} the decrease in observed flux from the afterglow
with time as for normal on-axis GRBs, and {\it ii)} an increase in flux
from the afterglow as more and more of the jet becomes visible to the 
observer with time. The combination of these effects results in a flatter 
early light curve than for GRB
afterglows and even a rising early light curve for some viewing angles
(see Granot et al.\ 2002; Rossi, Lazzati \& Rees 2002; and Dado et al.\ 
2003 for a discussion in the case of fireball and cannonball models 
respectively). In addition, the
shape of the jet may be important. In the collapsar model there is 
more slowly moving material further away from the jet axis, and the
early optical afterglow from an XRF will be dominated by emission from 
this material (Zhang, Woosley, \& Heger 2003;
Kouveliotou et al.\ 2004). The exact shape of the early light curve will 
hence be a function of the jet structure, the viewing angle, and the density 
profile of the environment. It is possible to reconcile the tentative
ROTSE-III detection with the off-axis scenario in the following way:
The steep decline following the constant phase indicate that this radiation
comes from material that is off-axis, i.e., our line of sight is outside the 
range of angles in which this material flows out. The decay rate during the 
ROTSE observations is poorly constrained, but a steep decay, approximately 
$t^{-2}$, is most consistent with our first data point. That means that this 
material, too, is viewed off-axis, and very likely is the same material that 
caused the off-axis prompt emission to which we attribute the XRF. Because the 
strength of off-axis emission decreases very steeply with angle away from the 
axis, this suggests that our line of sight is not too far outside the opening 
angle of the fast material that made the prompt emission. That then does not 
leave much solid angle for the slower material which caused the plateau. The
time of its break, about 1 day after trigger, implies an opening angle
of about 10 degrees for explosion energies and external densities
typical of ordinary GRBs, and a minimum initial Lorentz factor of 5--10
for this material. Its total energy must be larger than that of the
faster material, in order to cause the strong reflare responsible for the
flat light curve. It is important to note that for this interpretation it 
is not required that the slower
material be further off-axis than the fast material: it could also
occupy the same cone, but follow behind the fast material and cause a
re-energizing of the original forward shock.

We note that flat light curves have been seen also for a few
classical GRBs. The optical afterglow of GRB~970508 was almost
constant from 3~h to 24~h after the burst. It then rose
by about one magnitude before it entered the usual power-law decline 
(Pedersen et al.\ 1998). Panaitescu \& Kumar (2002) interpret this 
light curve as the result of a slightly off-axis GRB. The ROTSE-III
observations of the optical afterglow of GRB~030418 showed a rise
during the first 600~s and a roughly constant level during the 
following 1400~s (Smith et al.\ 2003b). Finally we note that the 
light curve of GRB~000301C displayed a flat light curve 3--4 days after 
the GRB (Jensen et al.\ 2001).

The late afterglow light curve, i.e.\ from 1 to 5--6 days after the burst, is 
also as expected for off-axis GRBs. The value $\alpha_2 \sim 2$ is the 
canonical late time slope (see e.g., Andersen et al.\ 2000, their Fig.~4.) 
reflecting the slope of the distribution of energies for relativistic
electrons producing the afterglow emission in the synchrotron model
(e.g., van Paradijs et al.\ 2000). Finally, the spectral slope of 
$\beta\approx1.0$ is also typical for GRB afterglows 
(e.g., Simon et al.\ 2001).

The afterglow of XRF~030723 
has also been detected in the X-ray band with the Chandra
X-ray telescope on July 25 and Aug.~4 (Butler et al.\ 2004). The properties 
of the X-ray afterglow is within the range of classical GRBs.
The decline of the X-ray afterglow from July 25 to Aug.~4 corresponds
to a power-law decay slope of $\alpha_X=1.0\pm0.1$ (Butler et al.\ 2004).
This is slower than the optical decay suggesting either a non-trivial
behaviour of the afterglow or that the bump could also
be present in the X-ray light curve. For a full discussion of X-ray
emission from GRBs, XRFs and SNe we refer the reader to Kouveliotou et al.\ 
(2004). 

\subsection{Limits on the Redshift 
}
\label{redshift}
We first discuss which limits we can place on the redshift from
the afterglow spectrum.
The lack of Ly$\alpha$ absorption down to 4000~\AA \, due 
to \ion{H}{1} in the intergalactic medium (the Ly$\alpha$ forest), 
allows us to infer a firm upper limit of $z\leq2.3$ for the redshift of 
XRF~030723. It should be noted that this is the first time a firm redshift 
limit can be set from an XRF afterglow spectrum - and that this has led
to the firm conclusion that this XRF is not simply a high redshift GRB.
If we would furthermore make the assumption that the absorption system
associated with the host galaxy of XRF~030723 has properties 
similar to those of GRB absorption systems, we can place 
stronger limits on the redshift. The 1$\sigma$ observed 
equivalent width ($W_{\mathrm{obs}}$) for absorption lines over the
region 4000--6400~\AA \ is about 1~\AA. This is the region
where the signal-to-noise ratio in the spectrum is highest and the 
spectrum is not affected by strong sky-subtraction residuals. GRB 
absorption systems appear to
always have strong Ly$\alpha$ absorption lines (Jensen et al.\ 2001;
Fynbo et al.\ 2001b; M\o ller et al.\ 2002; Hjorth et al.\ 2003b;
Vreeswijk et al.\ 2004). From the lack of Ly$\alpha$ absorption an upper limit 
of $z=2.1$ can be inferred. GRB absorbers also show \ion{C}{4} absorption 
lines at 1548/1550~\AA \ with rest 
equivalent width ($W_\mathrm{rest}$) 
in the range 1.2--5.1~\AA \ (see the same references as for Ly$\alpha$). 
Therefore a \ion{C}{4} line should have been detected at more than 
3$\sigma$ significance if the redshift is larger than $z=1.6$. Furthermore,
the \ion{Mg}{2} doublet at 2800~\AA \ has a total $W_\mathrm{rest}$ in the 
range 2.6--12~\AA \ for GRB absorption systems (e.g., Metzger et al.\ 1997;
Andersen et al.\ 1999; Vreeswijk et al.\ 2001; Jensen et al.\ 2001;
Castro et al.\ 2003; Jha et al.\ 2001). The absence of 
\ion{Mg}{2} absorption would exclude the redshift range $0.4 < z < 1.4$ 
under the assumption of absorption lines similar to GRB absorbers. 

\begin{figure*}[h]
\begin{flushleft}
{\includegraphics[width=2.00\columnwidth,angle=0,clip]{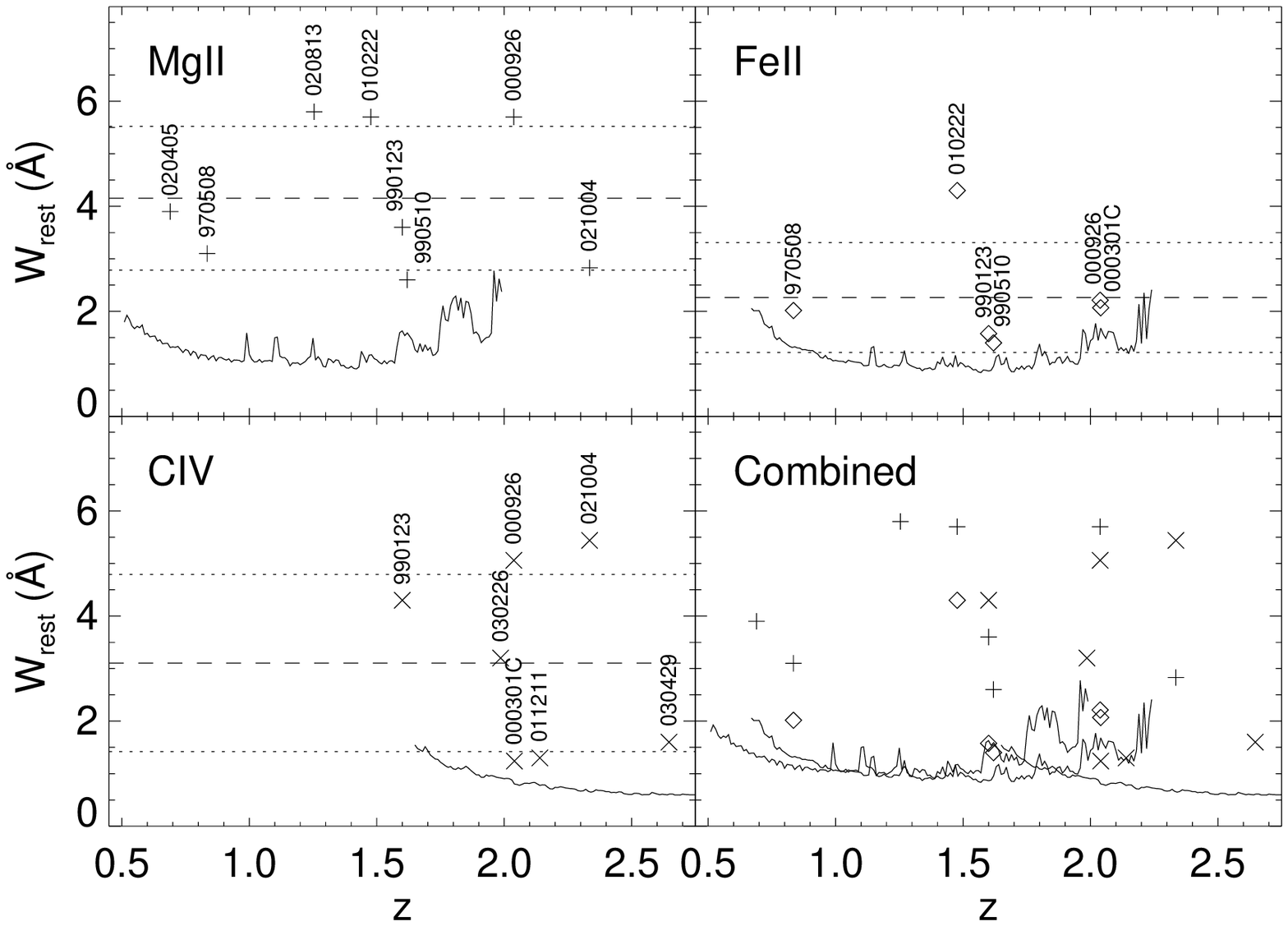}}
\caption{}{
Upper limit on the rest-frame equivalent width as a function of
redshift. In the three panels labelled \ion{Mg}{2}, \ion{Fe}{2}, and 
\ion{C}{4}
we present the $2\sigma$ upper limits (solid lines) on the total
$W_{\rm rest}$ measured in the afterglow spectrum as a function of assumed
redshift for the mentioned transitions. The upper limits are compared to
GRBs with known \ion{Mg}{2} (plotted with $+$), \ion{Fe}{2} (plotted with
$\diamond$) and \ion{C}{4} (plotted with $\times$) equivalent widths. The
dashed and dotted lines show the mean and $1\sigma$ spread of the $W_{\rm
rest}$ GRB measurements. The lower right panel shows the combined plot of
the three other panels.
}
\label{fig:zW}
\end{flushleft}
\end{figure*}

We have performed a cross-correlation between the observed afterglow 
spectrum and a simulated spectrum containing only absorption from 
\ion{Fe}{2} and \ion{Mg}{2}, but this
cross-correlation did not reveal any significant peaks. Due to the lack of
redshift information we turned to the following method. Assuming a redshift we
used the spectrum to measure the variance within one resolution element at
the expected position of a feature. By doing this for a sequence of
redshifts we can plot the $2\sigma$ upper limit on $W_{\rm rest}$
for \ion{Mg}{2}, \ion{Fe}{2}, and \ion{C}{4}. For \ion{Fe}{2} we only used the
two strongest features, \ion{Fe}{2} $\lambda 2382$ and $\lambda 2600$. The
upper limits are compared to known equivalent widths of GRBs in
Fig.~\ref{fig:zW}. The figure shows that if the absorption lines from 
the absorption system associated with XRF~030723 had been similar to the
absorbers associated with known GRBs, we would have expected an absorption
line from either \ion{Mg}{2}, \ion{Fe}{2}, or \ion{C}{4} if the redshift
is larger than $z=0.5$.

However, if at $z<0.5$ the host galaxy is extremely faint.
Bloom et al.\ (2003) argued that two XRF hosts were at $z>0.6$ based on their 
luminosities, and the host of XRF~030723 is almost 3 magnitudes fainter 
than these hosts. In fact, at $z<0.5$ the host galaxy has to be
fainter than $M\approx-15$. We therefore consider this low redshift unlikely.

It is more plausible that the absorption system associated
with XRF~030723 is weaker than typically found for GRB absorbers. This
would imply that XRF~030723 is located in an environment with lower
density than found for most GRBs so far. We consider the upper limit of 
$z < 2.3$ the only strong limit we have on the redshift of XRF~030723
based on the afterglow spectrum. 

Finally, we note that Atteia et al.\ (2004) have determined fiducial
redshifts (pseudo-z) for a sample of HETE bursts. The pseudo-z is based 
on the high-energy properties of the burst. For XRF~030723 Atteia et al.\
report a pseudo-z of 0.59.  

\subsection{The Nature of the Bump}
\label{bumpnature}
For long duration GRBs the relation to supernovae (SNe) became firmly 
established with the discovery of the type Ic supernova SN~2003dh 
associated with GRB~030329 (Stanek et al.\ 2003; Hjorth et al.\ 2003a). If
XRFs are simply the same phenomenon as long duration GRBs, but seen further
away from the jet axis, we would also expect to see supernovae underlying the 
light curves of XRFs. It would thus be natural to try to explain the bump
we see in Fig.~\ref{fig:lc} as the signature of a SN. 
The putative supernova peaks after about 16 days in the R-band, and the 
spectral energy distribution after the peak becomes
significantly redder than the early power-law spectra (see Sect.~\ref{SED}).
This motivates us to investigate whether the bump is consistent with
a supernova origin, in terms of peak luminosity, timing and SED.

The peak time of approximately 16 days after the XRF is early even for a
local supernova. The light curve peak is also unusually narrow. Fitting a
simple Gaussian to the peak gives a value of 1.7 mag for $\Delta$m$_{15}$,
i.e., the increase in magnitude between peak and 15 days later.
Correcting for time dilation corresponding to a
reasonable redshift, the intrinsic width gets even smaller (and
$\Delta$m$_{15}$ larger).
Early, narrow SN light curve peaks 
are expected from progenitor stars that have lost
(most of ) their hydrogen (and helium) envelopes. Of the known supernovae, 
the one
we have found to produce the best match to the light curve is the type Ic 
SN~1994I.
In the right panel of Fig.~\ref{fig:lc}
we plot the B-band light curve of SN~1994I redshifted to $z=0.6$ on top
of the light curve of XRF~030723. We have subtracted an extrapolation of
the power-law afterglow component (from the Beuermann fit). 
The errors include the photometric error as well as the formal error from the
subtraction of the Beuermann fit. As seen, the 
match is qualitatively acceptable except for data points before the peak. 
We cannot, however, exclude an afterglow light curve shape that is 
different from the purely empirical Beuermann function. Such a difference
would give large systematic errors on the data point prior to the peak.
In particular, a steeper late time decay slope would leave more room for 
a longer rise time for the SN.
SN~1994I displayed a very fast light curve (Richmond et 
al.\ 1996). Further, the observations of SN~1994I show that the peak
width becomes narrower at shorter wavelengths, so a near-UV light curve 
will surely be intrinsically more narrow than a B-band light curve, 
partially canceling the time dilation effect. In addition, the
decline rate, as measured by $\Delta m_{15}$, also differs
considerably between different type Ic supernovae. 
Richmond et al.\ (1996) estimate for SN~1994I that $\Delta$m$_{15}(U)
\sim 2.5$ (our estimate), $\Delta m_{15}(B) \sim 2.1$, and
$\Delta$m$_{15}(V) \sim 1.7$, while the peak magnitude was reached
after 8, 9 and 10 days ($\tau_\mathrm{peak}$), in the U, B and V bands,
respectively. For SN~1998bw we estimate the corresponding numbers from
Galama et al.\ (1998) to be $\Delta$m$_{15}(U) \sim 1.3$,
$\Delta$m$_{15}(B) \sim 1.0$, $\Delta$m$_{15}(V) \sim 0.7$, and
$\tau_\mathrm{peak}(U)= 13.5$ days, $\tau_\mathrm{peak}(B)= 14.7$ days, and
$\tau_\mathrm{peak}(V)= 17.0$ days. Finally,
SN 2002ap reached maximum brightness in U at 6.2 days, and in V at 9.9
days (Foley et al.\ 2003). There is thus a large variation in
both the rise time and decay time of type Ic supernovae. It is,
however, obvious that any supernova connected to XRF~030723 has to have
a light curve which is at least as fast as SN~1994I, and quite unlike 
SN~1998bw.

Although SN~1994I is the fastest type Ic to date, there is no reason
to exclude even faster scenarios. In a simple model, the width of the
light curve at the peak is set by $\tau_\mathrm{peak} \propto \kappa^{1/2}
M^{3/4} E^{-1/4}$ (e.g. Arnett 1996), where $E$ is the total energy,
$M$ the mass of the ejecta, and $\kappa$ the opacity. A faster light curve 
could therefore imply a smaller ejecta mass or/and a more
energetic explosion. A lower effective opacity, caused by e.g., a
higher effective temperature, would also decrease
$\tau_\mathrm{peak}$. Because the estimated C-O core mass of SN~1994I is only
$\sim 2.1 \Msun$ (Iwamoto et al.\ 1994), the most likely reason for a
faster time scale is a higher kinetic energy, possibly in
combination with a large asymmetry.  As several studies have shown
(e.g., Maeda \& Nomoto 2003; Woosley \& Heger 2003), a highly
asymmetric explosion may result in a fast rise time.  A less central
distribution of the radioactive $^{56}$Ni would further enhance this
effect. Because the diffusion time $\tau_{\rm d} \propto R^2 \rho
\kappa \propto 1/(V t)$, faster expansion in the polar direction leads
to a shorter $\tau_{\rm d}$ and a faster rise time (H\"oflich, Wheeler,
\& Wang 1999).

The absolute magnitude at the peak is determined by the amount of ejected
$^{56}$Ni, which is not necessarily correlated with the parameters
determining the width of the peak. The late bump of XRF 030723 peaked at
$\mathrm{R}\approx24.0$. 
This may provide a hint as to
the redshift for
the object. At a redshift of $z=0.4$, the apparent supernova bump would be
as faint as SN~1994I. While the putative supernova could in principle be
even fainter than the mediocre SN~1994I, even lower redshifts would also
make the host an unusually faint galaxy (see Sect.~\ref{redshift}).
On the other hand, for a redshift of $z=1.0$, the assumed bump would
correspond to a supernova about 
three times as bright as the powerful SN~1998bw.

There is not much information from the SED (Fig.~\ref{fig:sed}) to confirm a 
supernova origin for the late bump . 
However, the clear red evolution of the SED can exclude many other possible 
origins (see below).  
If the redshift is as high as $z\sim1$, most of the
observed optical emission (UBVRI) will correspond to the near-UV
regime, which is poorly known for type Ib/c supernovae at early
phases. An HST spectrum of SN~1994I does show a strong depression in the 
UV at 11 days past B maximum (Millard et
al.\ 1999). This is also seen in type Ia and type IIP supernovae, and
is interpreted 
as arising from the strong
UV blanketing due to the abundance
of overlapping lines from especially \ion{Ca}{2}, \ion{Fe}{1} and 
\ion{Fe}{2}, as well as
bound-free absorption by \ion{C}{2}. The only really early 
near-UV spectrum of a type Ic supernova we are aware of is a HST spectrum of 
SN~2002ap taken 6 days before maximum (courtesy of the SINS-team, 
P.I. Kirshner). This spectrum also shows a steep decline into the UV. 
The flux decreases by a factor ten between 4300 {\AA} to 2700 {\AA}. 

We stress, however,  
that a comparison with the very limited local sample of type
Ic supernovae may be misleading for several reasons. It is quite
likely that the X-ray flash emerges in the polar direction, where the
expansion velocities are highest and the density lowest. 
These parts of the ejecta may also have a higher $^{56}$Ni
fraction, and therefore a higher effective temperature. These factors may
together lead to a higher degree of ionization at these stages, which
would make UV line blanketing less effective. A measurable UV flux
at the time of the peak - as would be needed to interpret the light curve 
bump as due to a supernova at $z\sim1$,  could therefore result.

If we interpret the non-detections in the $UB$ bands after the peak as
due to the UV-blanketing, the redshift would have to be in the range
$z\sim0.3-1.0$. The late $K$ band detection also argues against a lower
redshift, to match the SED of known type Ic supernovae (e.g., Yoshii et
al.\ 2003).

The absence of any supernova features in the spectrum taken in the
bump-phase is not surprising since the expected small amplitude
undulations in this limited wavelength range are not likely to stand out
at this signal-to-noise ratio (see Fig.~\ref{fig:spec1}). This is true
for SNe with large expansion velocities like SN~1998bw. We can, however,
probably exclude SNe with lower expansion velocities, as for SN~1994I
(see Fig.~\ref{fig:spec1}). Very broad spectral features, and a narrow 
light curve, are expected from an energetic and asymmetric  explosion
viewed pole on (e.g., Iwamoto et al.\ 1998; H\"oflich et al.\ 1999).

In summary, an interpretation of the late, red light curve bump as due to a
supernova at redshift $z=0.3-1.0$ appears to be consistent with the
observations. This fits nicely
with the recent findings of a firm supernova connection for GRB 030329
(Garnavich et al.\ 2003; Stanek et al.\ 2003; Hjorth et al.\ 2003a) 
and with the spectroscopic redshift ($z=0.25$) of the probable 
host of XRF 020903 (Soderberg et al.\ 2003).
Regarding the SED we can only qualitatively say that the SED is consistent
with a supernova which is emitting strongly in the UV at maximum, but for
which the UV-depression sets in at later phases. A more secure and
quantitative statement is prevented by the poor knowledge of the early UV
spectra of type Ic supernovae and the lack of a spectroscopic redshift.

Other explanations to consider for the bump are {\it i)} a refreshed 
shock (Granot, Nakar, \& Piran 2003 and references therein), {\it ii)}
a two-jet model as the one proposed for GRB~030329 (Berger et al.\ 
2003; Huang et al.\ 2003), {\it iii)} the encounter with the 
termination shock of the progenitor wind or other density enhancements (e.g., 
Ramirez-Ruiz et al.\ 2001), {\it iv)} a dust-echo (Esin \& Blandford 2000; 
Reichart 2001), and  possibly {\it v)} microlensing 
(Garnavich, Loeb \& Stanek 2000). The sharp rise of the light curve in 
the bump 
phase is difficult to explain for density variations or refreshed shocks.
The light curve can be reasonably well described by a two-jet model
(Huang et al.\ 2003). However, the fundamental problem with most of the 
alternative models is to explain the SED. A power-law SED is expected at 
all epochs, and this is clearly excluded by the data. On the other hand, an
increasing absorption of blue light is probably a generic feature for a 
supernova. We find the dust-echo unlikely as the host appears to be
a tiny dwarf galaxy that most likely has a very low dust content. 
Therefore, we consider a SN the most likely explanation for the 
late light curve bump. Within this scenario, the evolution of the X-ray 
afterglow remains to be fully understood. 

\section{CONCLUSIONS}
We have presented the first detailed multi-color study of the afterglow
of an XRF. The properties of the early light curve are in agreement 
with the predictions for off-axis classical GRBs both in the fireball and
cannonball models. After the first week there is a strong bump in
the light curve, peaking at around 16 days after the XRF. The SED after the 
peak of the 
bump becomes significantly redder and can no longer be fit by a power-law. 
We have considered the possible interpretations of this bump, e.g., an 
underlying supernova, a second jet, a refreshed shock or an encounter with
a density enhancement. We conclude that a SN is the most likely explanation 
since no other model can qualitatively explain the 
evolution of the SED. We also present deep spectroscopy of the burst both 
in the afterglow and in the bump phase. Unfortunately, no redshift
determination has resulted from the spectra as there are no significant
absorption or emission lines detected. A firm upper limit of $z\leq2.3$ 
on the redshift is established from the lack of any Ly$\alpha$ forest lines
in the afterglow spectrum. If the bump in the light curve
is indeed due to a supernova the most likely redshift is in the range 
$\sim$0.3--1.0. The lack of significant absorption lines in the afterglow
spectrum and the lack of any indication of extinction in the afterglow
SED is indicative that the progenitor was in an environment with lower
density than typically found for GRBs. The host galaxy has not been
unambiguously detected. From the upper limit of $\mathrm{R}\gtrsim27.0$ and 
the preferred
redshift range of $z\approx0.3$--$1.0$ we conclude that the host is a 
dwarf galaxy.

\acknowledgments
We thank the anonymous referee for a thorough report that helped
us improve our manuscript.
This paper is based on observations collected by the Gamma-Ray Burst 
Collaboration at ESO (GRACE) at the European Southern Observatory, 
Paranal and La Silla, Chile. We thank the ESO staff for their help in 
securing the service mode data reported here. This publication makes 
use of data products from the Two Micron All Sky Survey, which is a 
joint project of the University of Massachusetts and the Infrared 
Processing and Analysis Center/California Institute of Technology, 
funded by the National Aeronautics and Space Administration and the 
National Science Foundation. The allocation of observing time by IJAF
at the Danish 1.54 m telescope on La Silla is acknowledged. A. J. 
Castro-Tirado aknowledges
the time devoted to the GRB target of opportunity programme
within the MOA Project at Mt. John. We acknowledge benefits 
from collaboration within the EU FP5 Research Training Network 
"Gamma-Ray Bursts: An Enigma and a Tool". This work was also 
supported by the Danish Natural Science Research Council (SNF) and 
by the Carlsberg Foundation.

\begin{deluxetable}{lccccc}
\tablecaption{Log of observations and photometry of the afterglow of XRF 030723.
Upper limits are measured in a circular aperture with diameter of 2 arcsec and 
are 2 sigma. 
\label{tbl:journal}}

\tablehead{
\colhead{Date} &
\colhead{$\Delta$t} &
\colhead{Filter/Grism} &
\colhead{Exposure time}&
\colhead{Instrument}&
\colhead{Brightness} \\
\colhead{(UT)}&
\colhead{(days)}&
\colhead{}&
\colhead{(s)} &
\colhead{}&
\colhead{(mag)} 
}
\startdata
July 23.44329 & 0.1736 & R    & $300$         & D1.5m/DFOSC & $20.92\pm0.11$ \\
July 23.6090  & 0.3394 & BV-MOA & $3\times300$ & 0.6m/MOA    & $21.47\pm0.25$ \\
July 23.6090  & 0.3394 & RI-MOA & $3\times300$ & 0.6m/MOA    & $>20.9$ \\
July 24.133   & 0.863  & R    & $3\times600$  & D1.5m/DFOSC & $20.98\pm0.05$ \\
July 24.333   & 1.063  & R    & $3\times600$  & D1.5m/DFOSC & $21.15\pm0.05$ \\
July 24.360   & 1.090  & V    & $3\times600$  & D1.5m/DFOSC & $21.63\pm0.03$ \\
July 24.384   & 1.114  & Js   & $15\times60$  & NTT/SOFI    & $19.81\pm0.06$ \\
July 24.397   & 1.127  & H    & $15\times60$  & NTT/SOFI    & $19.14\pm0.11$ \\
July 24.409   & 1.139  & Ks   & $15\times60$  & NTT/SOFI    & $18.33\pm0.06$ \\
July 24.421   & 1.151  & Js   & $15\times60$  & NTT/SOFI    & $19.79\pm0.08$ \\
July 24.433   & 1.163  & H    & $15\times60$  & NTT/SOFI    & $19.26\pm0.12$ \\
July 24.446   & 1.176  & Ks   & $15\times60$  & NTT/SOFI    & $18.42\pm0.10$ \\
July 24.396   & 1.126  & i    & $3\times600$  & D1.5m/DFOSC & $20.72\pm0.04$ \\
July 24.411   & 1.141  & R    & $4\times600$  & D1.5m/DFOSC & $21.24\pm0.05$ \\
July 25.114   & 1.844  & R    & $3\times600$  & D1.5m/DFOSC & $22.00\pm0.10$ \\
July 25.346   & 2.076  & R    & $7\times600$  & D1.5m/DFOSC & $22.31\pm0.07$ \\
July 25.378   & 2.108  & Js   & $15\times60$  & NTT/SOFI    & $20.14\pm0.16$ \\
July 25.391   & 2.121  & H    & $15\times60$  & NTT/SOFI    & $19.50\pm0.22$ \\
July 25.403   & 2.133  & Ks   & $15\times60$  & NTT/SOFI    & $19.08\pm0.25$ \\
July 25.408   & 2.138  & R    & $8\times600$  & D1.5m/DFOSC & $22.13\pm0.05$ \\
July 26.236   & 2.966  & V    & $50$          & VLT/FORS1   & $23.41\pm0.07$ \\
July 26.239   & 2.969  & V    & $50$          & VLT/FORS1   & $23.28\pm0.07$ \\
July 26.33    & 3.06   & G300V+GG375 & $22\times600$ & VLT/FORS1 & spectra \\
July 27.160   & 3.890  & Ks   & $10\times60$  & VLT/ISAAC   & $21.04\pm0.21$  \\
July 27.171   & 3.901  & Js   & $10\times60$  & VLT/ISAAC   & $22.46\pm0.23$  \\
July 27.180   & 3.910  & H    & $10\times60$  & VLT/ISAAC   & $21.06\pm0.15$  \\
July 27.212   & 3.942  & U    & $6\times300$  & VLT/FORS1   & $23.83\pm0.15$  \\
July 27.228   & 3.958  & B    & $300$         & VLT/FORS1   & $24.31\pm0.07$  \\
July 27.232   & 3.962  & B    & $300$         & VLT/FORS1   & $24.57\pm0.10$  \\
July 27.236   & 3.966  & V    & $300$         & VLT/FORS1   & $24.04\pm0.06$  \\
July 27.241   & 3.971  & V    & $300$         & VLT/FORS1   & $24.12\pm0.12$  \\
July 27.245   & 3.975  & R    & $300$         & VLT/FORS1   & $23.64\pm0.05$  \\
July 27.249   & 3.979  & R    & $300$         & VLT/FORS1   & $23.56\pm0.05$  \\
July 27.255   & 3.985  & i    & $300$         & VLT/FORS1   & $23.08\pm0.08$  \\
July 27.260   & 3.990  & i    & $300$         & VLT/FORS1   & $23.08\pm0.08$  \\
July 29.150   & 5.880  & R    & $300$         & VLT/FORS1   & $24.63\pm0.13$  \\
July 29.154   & 5.884  & R    & $300$         & VLT/FORS1   & $24.38\pm0.11$  \\
July 29.158   & 5.888  & R    & $300$         & VLT/FORS1   & $24.62\pm0.13$  \\
July 31.100   & 7.830  & R    & $6\times300$  & VLT/FORS2   & $25.00\pm0.14$  \\
Aug  3.240    & 10.970 & R    & $12\times300$ & VLT/FORS2   & $25.16\pm0.07$  \\
Aug  6.228    & 13.958 & V    & $6\times600$  & NTT/SUSI2   & $24.39\pm0.08$  \\
Aug  6.240    & 13.970 & i    & $6\times600$  & NTT/SUSI2   & $23.92\pm0.23$  \\
Aug  6.254    & 13.984 & R    & $6\times600$  & NTT/SUSI2   & $24.28\pm0.07$  \\
Aug  6.308    & 14.038 & R    & $6\times600$  & VLT/FORS1   & $24.14\pm0.03$  \\
Aug  6.368    & 14.098 & B    & $6\times600$  & NTT/SUSI2   & $25.04\pm0.10$  \\
Aug  7.291    & 15.021 & R    & $180$         & VLT/FORS2   & $24.16\pm0.09$  \\
Aug  7.296    & 15.026 & R    & $180$         & VLT/FORS2   & $24.20\pm0.11$  \\
Aug  7.359    & 15.089 & 600RI+GG435 & $15\times540$ & VLT/FORS2  & spectra  \\
Aug  8.327    & 16.057 & R    & $180$         & VLT/FORS2   & $24.04\pm0.13$  \\
Aug  8.331    & 16.061 & R    & $180$         & VLT/FORS2   & $24.03\pm0.10$  \\
Aug  8.380    & 16.110 & 600RI+GG435 & $10\times540$ & VLT/FORS2  & spectra  \\
Aug  9.396    & 17.126 & R    & $6\times300$  & VLT/FORS2   & $24.16\pm0.04$  \\
Aug 10.335    & 18.065 & R    & $14\times600$ & D1.5m/DFOSC & $23.88\pm0.44$ \\
Aug 14.168    & 21.898 & Ks   & $150\times60$ & VLT/ISAAC   & $21.19\pm0.11$  \\
Aug 17.052    & 24.782 & R    & $15\times600$ & NTT/SUSI2   & $25.08\pm0.09$  \\
Aug 19.007    & 26.737 & B    &  $3\times600$ & VLT/FORS1   & $>26.8$  \\
Aug 19.091    & 26.821 & U    & $15\times600$ & VLT/FORS1   & $>27.0$  \\
Aug 19.168    & 26.898 & R    & $3\times600$  & VLT/FORS1   & $24.99\pm0.06$  \\
Aug 19.191    & 26.921 & i    & $3\times600$  & VLT/FORS1   & $23.88\pm0.05$  \\
Sep  3.291    & 42.021 & R    & $6\times600$  & VLT/FORS1   & $25.97\pm0.12$ \\
Sep 25.135    & 63.865 & R    & $4\times600+480$ & VLT/FORS1 & $26.58\pm0.29$ \\
\enddata
\end{deluxetable}

\begin{deluxetable}{lccccc}
\tablecaption{
Evolution of the XRF 030723 SED
\label{tbl:sed}}

\tablehead{
\colhead{Observing Epoch} &
\colhead{Filters} &
\colhead{$\beta$}&
\colhead{$\chi^2/\mathrm{dof}$} \\
\colhead{(UT)}&
\colhead{}&
\colhead{}&
\colhead{}\\ 
}
\startdata
July 24.4 & VRiJ$_\mathrm{s}$HK$_\mathrm{s}$    & $0.96\pm0.04$   &   $1.5$  \\
July 25.4 & RJ$_\mathrm{s}$HK$_\mathrm{s}$      & $1.31\pm0.13$   &   $3.6$  \\
July 27.2 & UBVRiJ$_\mathrm{s}$HK$_\mathrm{s}$  & $1.00\pm0.07$   &   $2.2$  \\
Aug 6.3   & BVRi              & $1.03\pm0.25$   &   $3.1$  \\
Aug 19.0  & RiK$_\mathrm{s}$       & $1.15\pm0.10$   &   $58.3$ \\
\enddata
\end{deluxetable}

\end{document}